\documentclass[twocolumn,showpacs,prb,superscriptaddress,aps,floatfix]{revtex4-1}
\usepackage{rotating}
\usepackage{amsmath}
\usepackage{amsfonts}
\usepackage{amssymb}
\usepackage{epsfig}
\usepackage{float}

\usepackage{color}
\usepackage{graphicx}
\usepackage[active]{srcltx}
\usepackage[sort&compress]{natbib}

\newcommand{\qq}{{\bf q}}
\newcommand{\hh}{{\bf h}}
\newcommand{\rr}{{\bf r}}
\newcommand{\pp}{{\bf p}}
\newcommand{\kk}{{\bf k}}

\newcommand{\GG}{{\bf G}}
\newcommand{\SiS}{{\bf \Sigma}}
\newcommand{\VV}{{\bf V}}
\newcommand{\UU}{{\bf U}}

\newcommand{\be}{\begin{equation}}
\newcommand{\ee}{\end{equation}}
\newcommand{\ben}{\begin{equation*}}
\newcommand{\een}{\end{equation*}}
\newcommand{\bea}{\begin{eqnarray}}
\newcommand{\eea}{\end{eqnarray}}
\newcommand{\bean}{\begin{eqnarray*}}
\newcommand{\eean}{\end{eqnarray*}}

\renewcommand{\[}{\left[}
\renewcommand{\]}{\right]}
\renewcommand{\(}{\left(}
\renewcommand{\)}{\right)}

\newcommand{\grenoble}{Institut N\'eel,
CNRS/UJF, 25 rue des Martyrs BP 166, 38042 Grenoble cedex 9, France} 
\newcommand{\rome}{Dipartimento di Fisica, Universit\`a di Roma ``Tor
Vergata'', via della Ricerca Scientifica 1, I-00133 Roma, Italy}
\newcommand{\ikerb}{Ikerbasque, Basque Foundation for Science, E-48011 Bilbao, Spain}
\newcommand{\coimbra}{Centre for Computational Physics and Physics Department, University of Coimbra, Rua Larga, 3004-516 Coimbra, Portugal}
\newcommand{\etsf}{European Theoretical Spectroscopy Facility, NAPS/IMCN,
  Universit\'e Catholique de Louvain, B-1348 Louvain-la-Neuve, Belgium}

\begin{document}
\title{A real-time approach to the optical properties of solids and nano-structures: the time-dependent Bethe-Salpeter equation}

\author{C. Attaccalite}
\affiliation{\grenoble}

\author{M. Gr\"{u}ning} 
\affiliation{\coimbra}
\affiliation{\etsf}

\author{A. Marini}
\affiliation{\rome}
\affiliation{\ikerb}

\begin{abstract}
Many-body effects are known to play a crucial role in the electronic and optical properties of solids and nano-structures.
Nevertheless the majority of theoretical and numerical approaches able to capture the influence of Coulomb correlations 
are restricted to the linear response regime. In this work we introduce a novel approach based on a real-time solution of the electronic dynamics.
The proposed approach reduces to the well-known Bethe-Salpeter equation in the linear limit regime and it makes possible,
at the same time, to investigate correlation effects in nonlinear phenomena.
We show the flexibility and numerical stability of the proposed approach by calculating the dielectric constants and the effect of 
a strong pulse excitation in bulk {\it h}-BN.
\end{abstract}           


\maketitle

\section{Introduction}

Real-time methods have proven their utility in calculating optical properties of finite systems mainly within time-dependent density functional
theory (TDDFT).\cite{PhysRevB.62.7998,PSSB:PSSB200642067,*sun:234107} On the other hand extended systems have been mostly studied by using many-body  perturbation
theory (MBPT) within the linear response regime.~\cite{strinati} 
The different treatment of correlation and nonlinear effects mark the range of applicability of the two approaches. The real-time TDDFT makes
possible to investigate nonlinear effects like second harmonic generation\cite{takimoto:154114} or hyperpolarizabilities of 
molecular systems.\cite{PSSB:PSSB200642067} However the standard approaches used to approximate the exchange-correlation functional of TDDFT 
treat correlation effects only on a mean-field level. As a consequence, while finite systems---such as molecules---are well described, in the case of extended systems---such as periodic crystals and nano-structures---the real-time TDDFT does not capture the essential features of the optical absorption~\cite{RevModPhys.74.601} even qualitatively. 

On the contrary MBPT allows to include correlation effects using controllable and systematic approximations for the self-energy $\Sigma$,
that is a one-particle operator non-local in space and time.
$\Sigma$ can be evaluated within different approximations, among which one of the most successful is the so-called GW
approximation.\cite{Aulbur19991}
Since its first application to semiconductors\cite{PhysRevLett.45.290} the GW self-energy has been shown to
correctly reproduce quasi-particle energies and lifetimes for a wide range of materials.\cite{Aulbur19991}
Furthermore, by using the static limit of the GW self-energy as scattering potential of the
Bethe-Salpeter equation (BSE),\cite{strinati} it is possible to calculate response functions including electron-hole interaction
effects.

In recent years, the MBPT approach has been merged with density-functional theory (DFT) by using the Kohn-Sham Hamiltonian as zeroth-order term in the
perturbative expansion of the interacting Green's functions. This approach is parameter free and completely \emph{ab-initio},~\cite{RevModPhys.74.601} 
and in this paper will be addressed as {\it ab-initio}-MBPT ({\it Ai}-MBPT) to mark the difference with the conventional MBPT. 
However the {\it Ai}-MBPT is a very cumbersome technique that, based on a perturbative concept, increases its level of complexity with
the order of the expansion. As an example, this makes the extension of this approach beyond the linear response regime quite complex, though there have been recently some applications of the {\it Ai}-MBPT in nonlinear optics.~\cite{Chang2002,Leitsmann2005,PhysRevB.82.235201}

Another stringent restriction of the {\it Ai}-MBPT is that it cannot be applied when non-equilibrium phenomena take place: for example it cannot be applied to study the light emission after an ultra-fast laser pulse excitation.
A generalization of MBPT to non-equilibrium situations has been proposed by Kadanoff and Baym.\cite{kadanoffbaym} 
In their seminal works the authors derived a set of equations for the real-time Green's functions, the Kadanoff-Baym equations (KBE's), that provide the basic tools of the non-equilibrium Green's Function theory and allow essential advances in non equilibrium statistical mechanics.\cite{kadanoffbaym}   

Both the standard MBPT and non-equilibrium Green's Function theory are based on
the Green's function concept. This function describes the time propagation of a single particle excitation under the action of an external
perturbation.  
In the equilibrium MBPT, due to the time translation invariance,
the relevant variable used to calculate the Green's functions is the frequency $\omega$. Instead, out of equilibrium, in all non steady-state
situations, the time variables acquire a special role and much more attention is devoted to the their propagation properties. 
The time propagation avoids the explosive dependence, beyond the linear response, of the MBPT on high order Green's functions. Moreover the KBE are
non-perturbative in the external field therefore weak and strong fields can be treated on the same footing. 

One of the first attempts to apply the KBE's for investigating optical properties of semiconductors was presented in the seminal paper of Schmitt-Rink and co-workers.~\cite{PhysRevB.37.941} 
Later the KBE's were applied to study quantum wells,~\cite{PhysRevB.58.2064} laser excited
semiconductors,~\cite{PhysRevB.38.9759} and luminescence~\cite{PhysRevLett.86.2451}. However, only recently it was possible to simulate the Kadanoff-Baym
dynamics in real-time.~\cite{Kohler1999123,PhysRevLett.103.176404,PhysRevLett.84.1768,PhysRevLett.98.153004} 

In this work we combine a simplified version of the KBE's with DFT in such a way to obtain a parameter-free theory that is able to reproduce and predict
ultra-fast and nonlinear phenomena 
(Sec.~\ref{tdbse_section}). This approach, that we
will address as time-dependent BSE, reduces to the standard BSE for weak perturbations (Sec.~\ref{linear_response}) but, at the same time, 
naturally describes optical excitations beyond the linear regime. After discussing some relevant aspects of the practical implementation of our approach (Sec.~\ref{teospectro}), we 
exemplify how it works in practice by calculating the optical absorption spectra of {\it h}-BN  and the time dependent change in its electronic population due to 
the perturbation by means of an ultra-fast and ultra-strong laser pulse (Sec.~\ref{computational}).


\section{The time-dependent Bethe-Salpeter equation}
\label{tdbse_section}                                        
We derive here a novel approach to solve the time evolution of an electronic
system with Hamiltonian coupled with an external field,
\be
\label{hamiltonian}
\hat{H} = \hat{h} + \hat{H}_{mb} + \hat{U},     
\ee
where $U$ represents the electron-light interactions (see
Sec.~\ref{ss:solution} for its specific form). As usually done in MBPT, $\hat{H}$
is partitioned in an (effective) one-particle Hamiltonian
$\hat h$ and a part containing the many-particle effects $\hat{H}_{mb}$. 

In our derivation, we take as starting point the KBE's that we briefly introduce in
Sec.~\ref{ss:KBEs} (see e.g. Refs.~\onlinecite{kremp} for a systematic
treatment). Then, in Sec.~\ref{ss:KS-CHSX} we proceed in analogy with
the equilibrium \emph{Ai}-MBPT: first, we define $\hat h$ as the Hamiltonian of the Kohn-Sham
system, second we introduce the same approximations for the
self-energy operator. As a result we obtain the analogous of the
successful $GW$+BSE approach for the non-equilibrium case. Indeed in
Sec.~\ref{linear_response} we show that our approach, the
time-dependent BSE, reduce to the $GW$+BSE in the linear regime.

\subsection{The Kadanoff-Baym equations}
\label{ss:KBEs}
Within the KBE's, the time evolution of an electronic system 
coupled with an external field is described by the equation of motion for the non-equilibrium Green's functions\cite{kadanoffbaym,kremp,schafer}, $G\(\rr,t;\rr' t'\)$.  
To keep the formulation as simple as possible and, being interested
only in long wavelength perturbations,
we expand the generic $G$ in the eigenstates $\{\varphi_{n,\kk}\}$ of the $\hat{h}$ Hamiltonian for a fixed momentum point $\kk$:
\begin{multline}\label{GFdefinition}
\[\GG_{\kk}\(t_1,t_2\)\]_{n_1 n_2}\equiv G_{n_1 n_2, \kk }(t_1,t_2)=\\\int
\varphi^*_{ n_1\kk}({\mathbf r}_1) G\(\rr_1,t_1;\rr_2,t_2\) \varphi_{n_2\kk}({\mathbf
r}_2){\mathrm d}^3r_1{\mathrm d}^3r_2.
\end{multline}
As the external field does not break the spatial invariance of the system $\kk$ is conserved.

Within a second-quantization formulation of the many-body problem, the equation of motion for the Green's function described by Eq.~\eqref{GFdefinition} are obtained
from those for the creation and destruction operators.  However the resulting equations of motion for $\GG_{\kk}$ are not closed:
they depend on the equations of the two-particle Green's function that in turns depends on the three-particle Green's function and so on.
In order to truncate this hierarchy of equations, one introduces the self-energy operator $\SiS_{\kk}(t_1,t_2)$, a non-local and frequency dependent
one-particle operator that holds information of all higher order Green's functions. 
A further complication arises with respect to the equilibrium case because of the lack of time-traslation invariance in non-equilibrium phenomena that implies that 
$\SiS_{\kk}(t_1,t_2)$ and $\GG_{\kk}(t_1,t_2)$ depend explicitly on both $t_1,t_2$. Then, one can define an
advanced $\SiS_{\kk}^a$ ($\GG_{\kk}^{\mathrm a}$), a retarded $\SiS_{\kk}^r$ ($\GG_{\kk}^{\rr}$), a greater and a lesser $\SiS_{\kk}^>,\SiS_{\kk}^<$ ($\GG_{\kk}^>,\GG_{\kk}^<$) self-energy
operators (Green's functions) depending on the ordering of $t_1,t_2$ on the time axis. 
Finally, 
the following equation for the $\GG_{\kk}^<$ is obtained (see e.g. Ch.~2 of
Ref.~\onlinecite{kremp} for more details):
\begin{multline}
 i\hbar  \frac{\partial}{\partial t_1} G^<_{n_1n_2\kk}(t_1,t_2)=\mbox{}  \delta(t_1-t_2)\delta_{n_1n_2}    \\
+  h_{n_1n_1\kk}(t_1) G^<_{n_1n_2\kk}(t_1,t_2) + \sum_{n_3}U_{n_1n_3\kk}(t_1) G^<_{n_3n_2\kk}(t_1,t_2) \\
+  \sum_{n_3} \int \mathrm{d}t_3 \big( \Sigma^{\rr}_{n_1n_3\kk}(t_1,t_3)G^<_{n_3n_2\kk }(t_3,t_2) \\
+ \Sigma^<_{n_1n_3\kk}(t_1,t_3) G^{\mathrm a}_{n_3n_2\kk}(t_3,t_2)\big) .
\label{eqmotGone}
\end{multline}
This equation, together with the adjoint one for $ i\hbar  \frac{\partial}{\partial t_2} G^<$, describes the
evolution of the lesser Green's function $\GG_{\kk}^<$ that gives
access to the electron distribution ($\GG_{\kk}^<(t,t)$) and to the
average of any one-particle operator such as for example the electron
density [Eq.~\eqref{eqden}], the polarization [Eq.~\eqref{polarization}] and the current.  However, in general $\Sigma^{\rr},\Sigma^<$ and the $\GG_{\kk}^{\mathrm a}$ depend on $\GG_{\kk}^>$, so that in addition to Eq.~\eqref{eqmotGone} the corresponding equation for the $\GG_{\kk}^>$ has to be solved. 

Then, in principle, to determine the non-equilibrium Green's function in presence of an external perturbation one needs  to solve the system of coupled equations for $\GG_{\kk}^>,\GG_{\kk}^<$, known as KBE's. Indeed, this system has been implemented within several approximations for the self-energy in model
systems,\cite{Kohler1999123,PhysRevLett.103.176404} in the homogeneous electron gas,\cite{PhysRevLett.84.1768} and in atoms\cite{PhysRevLett.98.153004}.
The possibility of a direct propagation in time of the KBE's provided, in these systems, valuable insights on the real-time dynamics of the electronic excitations, as their lifetime and transient effects.\cite{Kohler1999123,PhysRevLett.103.176404,PhysRevLett.84.1768,PhysRevLett.98.153004} 
Nevertheless, the enormous computational load connected to the large number of degrees of freedom {\it de facto} prevented the application of this method to crystalline solids, large molecules and nano-structures. In the next subsection we show a simplified approach---grounded on the DFT---that while capturing most of the physical effects we are interest in, makes calculation of ``real-world'' systems feasible. 

\subsection{The Kohn-Sham Hamiltonian and an approximation for the self-energy}
\label{ss:KS-CHSX}

In analogy to {\it Ai}-MBPT for the equilibrium case,
we choose as $\hat{h}$ in Eq.~\eqref{hamiltonian} the Kohn-Sham Hamiltonian,~\cite{PhysRev.140.A1133} 
\be
\hat{h} = -\frac{\hbar^2}{2m} \sum_i \nabla_i^2 + 
\hat{V}_{eI} + \hat{V}^H[\tilde \rho]   + \hat{V}^{xc}[\tilde \rho], \label{eq:kshamH}
\ee
where $\hat{V}_{eI}$ is the electron-ion interaction, 
$\hat{V}^H$ is the Hartree potential and
$\hat{V}^{xc}$ the exchange-correlation potential.
Within DFT, the Kohn-Sham Hamiltonian corresponds to the independent particle system
that reproduces the ground-state electronic density $\tilde \rho$ of
the full interacting system ($\hat h + \hat H_{\text{mb}}$), that is
\be
\tilde \rho = \sum_{n\kk} f_{n\kk}|\varphi(\rr)|^2,
\label{eq:KSden}
\ee
where $f_{n\kk}$ is the Kohn-Sham Fermi distribution. 

Equation~\eqref{eqmotGone} can be greatly simplified by choosing 
a static retarded approximation for the self-energy,
\begin{subequations}
\begin{align}
{\bf\Sigma}^{\rr} (t_1,t_2) & = \left[ {\bf\Sigma}^{\mathrm {cohsex}} (t_1) - {\bf V}_{xc} \right ] \delta(t_1 - t_2) \label{eq:appcsx}\\
{\bf\Sigma}^<(t_1,t_2) & =0 \label{eq:appret}
\end{align}
\end{subequations}
where the usual choice is ${\bf\Sigma}^{\mathrm {cohsex}}$, the so-called Coulomb-hole plus screened-exchange
self-energy (COHSEX). In Eq.~\ref{eq:appcsx} we subtracted the correlation effects already accounted by Kohn-Sham Hamiltonian $\hat{h}$.\cite{PhysRevB.38.7530}
The COHSEX is composed of two parts:
\bea
\Sigma^{\text{sex}}(\mathbf r,\mathbf r',t)=i W(\mathbf r,\mathbf r'; G^<)G^<(\mathbf r,\mathbf r',t), \\
\Sigma^{\text{coh}}(\mathbf r,\mathbf r',t)= -W(\mathbf r,\mathbf r'; G^<) \frac{1}{2}\delta(\mathbf r-\mathbf r'),
\label{coh_anx_sex}
\eea
where $W(\mathbf r,\mathbf{r'}; G^<)$ is the Coulomb interaction in
the random-phase approximation (RPA). These two terms are obtained as a
static limit of the $GW$ self-energy (see Ch.4 of Ref.~\onlinecite{kremp} and
Refs.~\onlinecite{PhysRevB.38.7530, PhysRevB.69.205204}). 

With the approximation in Eqs.~\eqref{eq:appcsx}--\eqref{eq:appret}, Eq.~\eqref{eqmotGone} does not depend anymore
on ${\bf G}^>$ and it is diagonal in time:
\begin{multline}
\label{eqmotGcohsex}
 i\hbar  \frac{\partial}{\partial t} G_{n_1,n_2,\kk}^<(t)=\\
 \left [ \hh_{\kk} + \UU_{\kk}(t) +  \VV_{\kk}^H[\rho] -
   \VV_{\kk}^H[\tilde \rho] \right .\\
   \left . + (\SiS_{\kk}^{\mathrm{cohsex}} (t)
   -\VV_{\kk}^{xc}[\tilde \rho]), \GG_{\kk}^<(t)\right ]_{n_1,n_2}.
\end{multline}
where $\rho$ is the density obtained from the $G^<$ as 
\be
\label{eqden}
\rho(\mathbf r,t) = \frac{i}{\hbar} \sum_{n_1n_2\kk} \varphi_{n_1 \kk} (\mathbf r)  \varphi^*_{n_2 \kk}  (\mathbf r) G^<_{n_2n_1\kk}(t).
\ee
 
Equation~\eqref{eqmotGcohsex} is conserving~\cite{kadanoffbaym} and satisfies the
sum rules for the response functions because both the one- and
two-particle self-energies are obtained from the same
$\Sigma^{\mathrm{cohsex}}$ 
and the system of equations is solved self-consistently.~\cite{refId}

%
However, despite the full real-time COHSEX dynamics [Eq.~\ref{eqmotGcohsex}] is an appealing
option considerably simplifying the dynamics with respect to the KBE's,
it neglects the dynamical dependence of the self--energy operator. This, in practice, induces a consistent renormalization 
of the quasiparticle charge\cite{PhysRevLett.45.290} in addition to an opposite enhancement of the optical properties~\cite{PhysRevLett.91.176402}.
In the COHSEX approximation both effects are neglected.
At the level of response properties for
most of the extended systems dynamical effects are either negligible or very small
(while recently it has been shown their importance for finite
systems, see Refs.~\onlinecite{bsedynamic,PhysRevB.77.115118}) and, for practical  purposes, it has been shown that they  
partially cancel with the  quasi-particle renormalization factors.~\cite{PhysRevLett.91.176402} 

Therefore we modify Eq.~\ref{eqmotGcohsex} in order to include only the effect
of the dynamical self-energy on the renormalization of the quasi-particle energies, that is the most
important effect.
Also in this case, our idea is to proceed in strict analogy with {\it
  Ai}-MBPT and to derive a real-time equation that reproduces the fruitful combination of the $G_0W_0$
approximation---for the one-particle Green's function---and of the BSE with
a static self-energy---for the two-particle Green's function. 
Indeed the $G_0W_0$+BSE is the state-of-the-art approach to study optical
properties within the {\it Ai}-MBPT.~\cite{RevModPhys.74.601} 
To this purpose Eq.~\eqref{coh_anx_sex} is modified as:
\begin{multline}
\label{tdbse}
 i \hbar  \frac{\partial}{\partial t} G^<_{n_1n_2\kk}(t)= \\
\left [ \hh_{\kk} + \Delta \hh_{\kk} + \UU_{\kk} +\VV_{\kk}^H[\rho] - \VV_{\kk}^H[\tilde \rho] \right . \\
 \left . + \SiS_{\kk}^{\text {cohsex}}[G^<] - \SiS_{\kk}^{\text{cohsex}}[\tilde G^<], \GG_{\kk}^<\(t\) \right ]_{n_1n_2}. 
\end{multline}
$\Delta \hh$ is a scissor operator\cite{RevModPhys.74.601} that
applies the $G_0 W_0$ correction to the Kohn-Sham eigenvalues, $e_{n_1\kk}^{KS}$,
\be
\label{eq:sciss}
\[\Delta \hh_{\kk}\]_{n_1,n_2} =  \left ( e_{n_1\kk}^{G_0W_0}  -e_{n_1\kk}^{KS}  \right ) \delta_{n_1,n_2},
\ee
and $\tilde G^<_{n n'}$ is the solution of Eq.~(\ref{tdbse}) for the
unperturbed system ($U=0$)
\be
\tilde G^<_{n n' \kk} = i \hbar f_{n\kk} \delta_{n n'}, \label{gtilde}
\ee
where we assume that the Kohn-Sham Fermi distribution is not changed by the
scissor operator. Note further that in Eq.~\ref{tdbse}, $V^{xc}[\tilde \rho]$ cancels out because it is independent of $G^<(t)$.

Equation~\eqref{tdbse} is the key result of this work. It is equivalent to assume that the
quasi-particle corrections modify only the single particle eigenvalues
leaving unchanged the Kohn-Sham wave functions.  Within {\it Ai}MBPT this approximation
is very successful for a wide range of materials characterized by weak correlations (see e.g. Refs.~\onlinecite{RevModPhys.74.601,Aulbur19991}). 

\begin{figure}[t]
\centering
\epsfig{figure=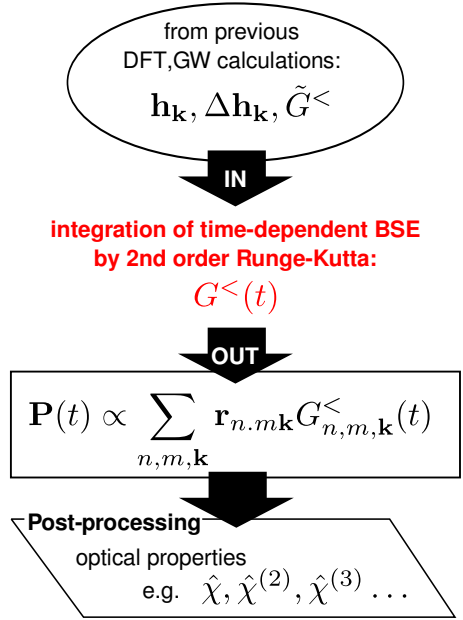,width=6cm}
\caption{\footnotesize{
Schematic flow of a time-Dependent BSE simulation. See
Sec.~\ref{ss:solution} for details.}} \label{fg:scheme}
\end{figure}

\subsection{The linear response limit}
\label{linear_response}

When an external perturbation $U(t)$ is switched on in Eq.~\eqref{tdbse}, it induces a variation of the Green's function, 
$\Delta \GG_{\kk}^<(t) = \GG_{\kk}^<(t) -  \tilde{\GG}_{\kk}^<$.
In turns, this variation induces a change in the 
self-energy and in the Hartree potential.
In the case of a strong applied laser field these changes depend on
all possible orders in the external field. However for weak fields the
linear term is dominant. 
In this regime it is possible to show analytically that Eq.~\eqref{tdbse}  reduces to the $G_0W_0$+BSE approach\cite{strinati,Aulbur19991}.
Proceeding similarly to Ref.~\onlinecite{bsedynamic} we consider the retarded density-density correlation function:
\begin{multline}
\label{chi-rr}
\chi^{\mathrm{r}}(\rr,t;\rr',t') =
-i\[\langle \rho(\rr,t)\rho(\rr',t')\rangle \right. \\ \left.
- \langle \rho(\rr,t)\rangle \langle \rho(\rr',t')\rangle\]\theta\(t-t'\).
\end{multline}
$\chi^{\mathrm{r}}$ describes the linear response of the system to a weak
perturbation, represented in Eq.~\eqref{hamiltonian}  by $U$,
\be
\label{eq:phychi}
\chi^{\mathrm{r}}(\rr,t;\rr',t') = \left. \frac{\langle
  \delta\rho(\rr t) \rangle}{\delta U(\rr^\prime t^{\prime})}\right\vert_{U=0}.
\ee
We start by expanding  $\chi (\mathrm{r})$ in terms of the Kohn-Sham orbitals:
\begin{multline}
\label{basisChi}
\chi^{\mathrm r}(\rr,t; \mathbf{r'},t'; \mathbf q) =
  \sum_{\substack{i,j,\kk \\ l,m,\kk^\prime} } \chi^{\mathrm r}_{\substack{i,j,\kk \\ l,m,\kk^\prime} }(t,t^\prime; \mathbf q)\\
\times \varphi_{i,\kk} (\rr)\varphi^*_{j, \mathbf {k+q}} (\rr) \varphi^*_{l, \mathbf {k^\prime}} (\rr')
\varphi_{m, \mathbf {k^\prime+q}} (\rr'),
\end{multline}
where $\qq$ is the momentum, and we define the matrix elements of $\chi^{\mathrm r}$ as,
\begin{multline}
\label{eq:chimat}
 \chi^{\mathrm r}_{\substack{ij,\kk \\ lm,\kk^\prime} }(t,t^\prime; \mathbf q) =\\
\iint \mathbf{d}^3r \mathbf{d}^3r^\prime \varphi^*_{i,\kk}(\rr)
\varphi^*_{m, \kk^\prime+\qq} (\rr^\prime)
\varphi_{j,\kk+\qq} (\rr) \varphi_{l, \kk^\prime} (\rr^\prime).
\end{multline}
Since we are interested only in the optical response, in what follows
we restrict ourselves to the case $\qq =0$ and drop the $\qq$
dependence of $\chi^{\mathrm r}$ (for the extension to finite momentum
transfer see Ref.~\onlinecite{PhysRevLett.84.1768}). 
Inserting the expansion for $\chi$ [Eq.~\eqref{basisChi}], $\rho$ [Eq.~\eqref{eqden}] and $U$ ($U_{m,n \kk}\equiv \langle m \kk | U | n \kk \rangle$) in Eq.~\eqref{eq:phychi} we obtain the following relation linking the matrix elements of $\chi^{\mathrm r}$ to the matrix elements of $G^<$:
\be
\label{DDcorFun_n}
\chi^{\mathrm r}_{\substack{ij,\kk \\ lm, \pp }}(t,t^\prime) =
\left. \frac{\delta\langle i G^<_{ji,\kk}(t)\rangle}{\delta U_{lm,\pp}(t^{\prime}
  )}\right\vert_{U=0}.
\ee
Then, we can obtain the equation of motion  for the matrix elements of $\chi^{\mathrm r}$
by taking the functional derivative of Eq.~\eqref{tdbse} with respect to $U_{l,m,\kk}\(t\)$,
\begin{multline}
\label{dtGtt}
-i\hbar \frac{\partial}{\partial t} \chi^{\mathrm r}_{\substack{ij, \kk \\ lm,\pp }}(t,t^\prime)= \\ 
\frac{\delta}{ \delta U_{lm, \pp}(t^{\prime})} [\hh_{\kk} + \Delta \hh_{\kk} + \UU_{\kk}(t) +  \VV_{\kk}^H[\rho(t)] -\VV_{\kk}^H[\tilde \rho]  \\
+  \SiS_{\kk}[G^<(t)] - \SiS_{\kk}[\tilde G^<], \GG_{\kk}^<(t) ]_{\substack{ji}}. 
\end{multline}
Making use of the definitions in Eqs.~\eqref{eq:sciss} and \eqref{gtilde}, together with Eq.~\eqref{DDcorFun_n}, it can be verified that the functional derivative of the one-electron Hamiltonian and 
of the external field give the contribution 
\begin{multline}
\label{diag_contr}
\left. \frac{\delta}{ \delta U_{l,m, \pp}(t^{\prime})} [\hh_{\kk}+\Delta \hh_{\kk} + \UU_{\kk},\GG_{\kk}^<(t)]_{\substack{ji}} \right\vert_{U=0}=\\
(e^{G_0W_0}_{j\kk} - e^{G_0W_0}_{i\kk})  \chi^{\mathrm r}_{\substack{ji, \kk \\ lm,\pp }}(t-t^\prime) + i(f_{i\kk}-f_{j\kk})\delta_{jl}\delta_{im}\delta_{\kk\pp}. 
\end{multline}
Note that, since the perturbation is weak, the Hamiltonian of the system is invariant with respect to time translation and thus $\chi^{\mathrm r}$ depends only on $ t-t^\prime$.
The term in Eq.~\eqref{dtGtt} containing the Hartree potential, that is not directly depending on the  external perturbation, is expanded  with respect to $U_{l,m,\kk}\(t\)$ by using  the functional derivative chain rule and the definition in Eq.~\eqref{DDcorFun_n} as: 
\begin{multline}\label{V_exp}
\delta V^H_{ij,\kk}\[\rho\(t\)\]= 
 \sum_{\substack{n,n',\pp\\l,m,\kk^\prime}} \iint\,dt^\prime\,dt^{\prime\prime} \frac{\delta V^H_{ij,\kk}\[\rho\(t\)\]}{\delta G^<_{n'n,\pp}\(t^\prime\)}\\ \times \chi^{\mathrm r}_{\substack{n,n',\pp\\lm,\kk^\prime}}(t^\prime,t^{\prime\prime})\delta U_{lm,\kk^\prime}\(t^{\prime\prime}\),
\end{multline}
A similar equation can be obtained for $\Sigma^{\text{cohsex}}_{ij,\kk}[G^<(t)]$.
Equation~\eqref{V_exp} for Hartree potential and its analogous for the self-energy can be explicited by using 
\begin{align}
V^H_{mn,\kk}(t)=&-2i \sum_{ij} G^<_{ji,\kk}\(t\) v^{\qq=0}_{\substack{mn, \kk \\ ij,\kk }}, \label{eq:hrtrG}\\
\Sigma^{\text{cohsex}}_{mn,\kk}(t)=& i \sum_{ij,\qq} G^<_{ji,(\kk-\qq)}\(t\) W_{\substack{m \kk,i (\kk-\qq) \\ n\kk,j (\kk-\qq) }},\label{eq:chsxG}
\end{align}
where the matrix elements of  $v^{\qq=0}$ and $W$ are labeled accordingly to Eq.~\eqref{eq:chimat}.
In Eq.~\eqref{eq:hrtrG} $v^{\qq=0}$ is the long range part of the bare Coulomb potential, responsible for the local field effects in the
BSE.   
Then by inserting Eq.~\eqref{eq:hrtrG} in Eq.~\eqref{V_exp} the functional derivative for the Hartree term is
\begin{multline}
\label{H_contr}
\left. \frac{\delta}{ \delta U_{lm, \pp}(t^{\prime})} \left[\VV_{\kk}^H[\rho(t)] -\VV_{\kk}^H[\tilde \rho],\GG_{\kk}^<(t)\right]_{\substack{ji}} \right\vert_{U=0}=\\
\(2i^2\)\(f_{i\kk} - f_{j\kk}\) \sum_{st} v^{\qq=0}_{\substack{ji, \kk \\ st,\kk }} \chi^{\mathrm r}_{\substack{st, \kk \\ lm,\pp }}(t-t^\prime).
\end{multline}

\begin{figure*}[t]
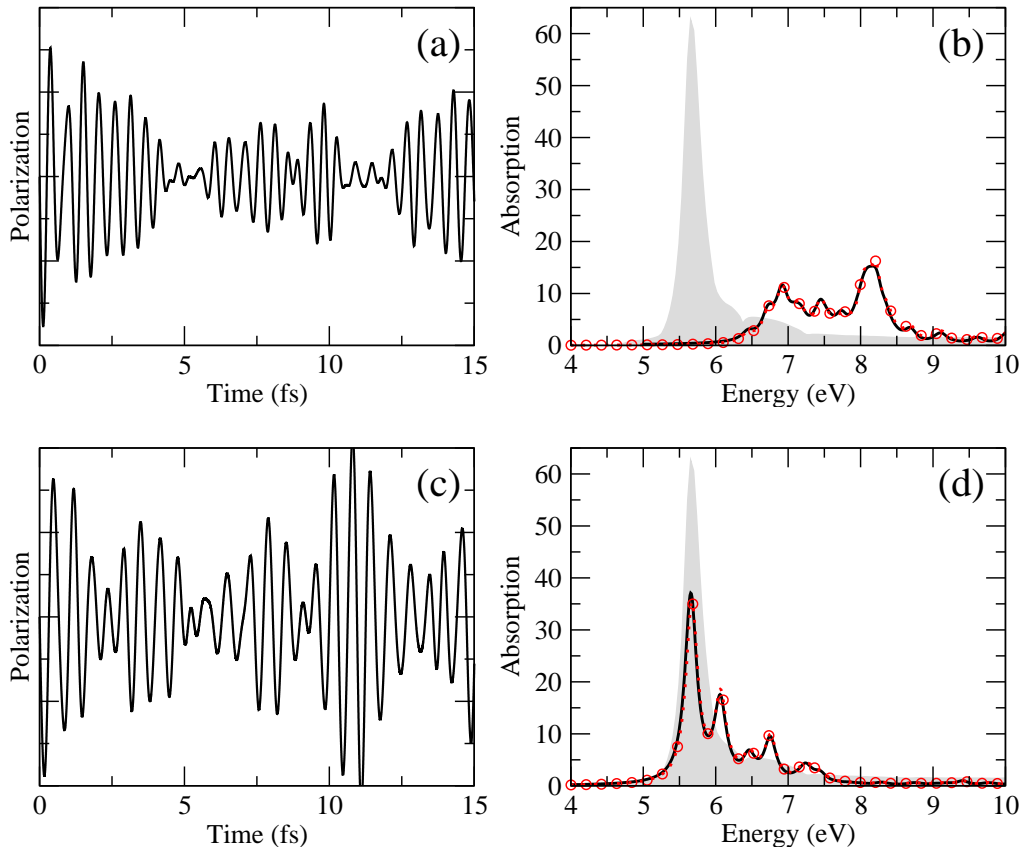

\centering
\epsfig{figure=tdh.eps, clip=,bbllx=20,bblly=260,bburx=715,bbury=530,width=15cm}
\epsfig{figure=bse.eps, clip=,bbllx=20,bblly=260,bburx=715,bbury=530,width=15cm}
\caption{\footnotesize{
{\it h}-BN: Comparison between the real-time approach and the
standard RPA and BSE approaches based on the equilibrium MBPT. {\bf(a)},{\bf(c)}: polarization $\mathbf P(t)$ generated by an electric field
$\mathbf E(t)=\mathbf E_o \delta(t)$ within the TD-HARTREE [{\bf(a)}] and
TD-BSE [{\bf(c)}] approximations.{\bf(b)},{\bf(d)}: the corresponding
absorption spectra (red circles) are compared with the RPA [{\bf(b)}]
and with the BSE [{\bf(d)}] results (black line). The experimental
absorption spectrum (grey shadow) is also shown as reference.}}
\label{hbn}
\end{figure*}

Similarly, an analogous equation is obtained for the self-energy (see also Appendix \ref{fastcohsex}), 
\begin{multline}
\label{xc_contr}
\left. \frac{\delta}{ \delta U_{lm, \pp}(t^{\prime})} \left[\SiS_{\kk}[\GG^<(t)] -\SiS_{\kk}[\GG^<(t)],\GG_{\kk}^<(t) \right]_{\substack{ji}}\right\vert_{U=0}=\\
\(-i^2\)\(f_{i\kk} - f_{j\kk}\) \sum_{st,\qq}  W_{\substack{j \kk,s (\kk-\qq) \\ i\kk,t (\kk-\qq) }} \chi^{\mathrm r}_{\substack{st,(\kk-\qq) \\ lm,\pp }}(t-t^\prime),
\end{multline}
where we neglected the part containing the functional derivative of the screened interaction with respect to the external perturbation. This is a basic assumption of the standard BSE that is introduced in order to neglect high order vertex corrections.\cite{strinati}

Finally, we insert Eqs.~\eqref{diag_contr},~\eqref{H_contr} and~\eqref{xc_contr} in Eq.~\eqref{dtGtt}, and by Fourier transforming with respect to $(t-t^\prime)$
we obtain
\begin{multline}
\label{bse}
 \left[  \hbar \omega- \left(\epsilon_{j \kk}^{\mathrm{G_0W_0}} -\epsilon_{i \kk}^{\mathrm{G_0W_0}}\right) \right]
\chi^{\mathrm r}_{\substack{ij,\kk\\lm \mathbf p}}(\omega) = \\
i \(f_{i \kk}-f_{j \kk}\) \left[ \delta_{jl} \delta_{im} \delta_{\kk,\pp} + \right. \\ \left.
+i\sum_{st,\qq}\{ W_{\substack{j \kk,s (\kk-\qq) \\ i\kk,t (\kk-\qq) }} -2   v^{\qq=0}_{\substack{ji, \kk \\ st,\kk }}  \}  
\chi^{\mathrm r}_{\substack{st, \kk-\qq \\ lm,\pp }}\(\omega\) \].
\end{multline}
formally equivalent to the standard BSE. 


\section{Optical properties from a time-dependent approach}
\label{teospectro}      
\subsection{Practical solution of the time-dependent BSE}
\label{ss:solution}
To solve Eq.~\eqref{tdbse} for a given electronic system [Eq.~\eqref{hamiltonian}], we start from $\hat{h}$, with its eigenvalues and eigenstates determined from a previous DFT calculations, and from the corrections $\Delta \hh_\kk$, determined e.g. from a previous $G_0W_0$ calculation.
Then, we switch on the external perturbation $U$ and integrate the
equations of motion using the same scheme as in Ref.~\onlinecite{Kohler1999123} for the diagonal part of the $G^<$, that is equivalent to a second order Runge-Kutta. Specifically, in Eq.~\eqref{hamiltonian} we choose to treat the interaction with the external electric field $\mathbf E$ within the direct coupling---or length gauge,
\be
\hat{U} = - e\hat{\mathbf r} \cdot \mathbf E(t). \label{eq:lengau}
\ee
Other choices are possible and indeed in the literature the electron-light interaction is often described within the
minimal coupling---or velocity gauge $( \hat{\mathbf p} \cdot \mathbf A)$, with $\mathbf A$ the vector potential.
As it has been pointed out in
Ref.~\onlinecite{boyd_gauge,*PhysRevA.36.2763}, the length and
velocity gauges lead to the same results only if a gauge
transformation is correctly applied.  However, in this respect the
velocity gauge presents two main drawbacks. First, the wave functions and the boundary conditions have to be transformed by a time-dependent gauge factor $T(\mathbf r,t) = \exp \{ i \mathbf A(t) \cdot \hat{\mathbf r}\}$
and accordingly, in the Green's function formalism also the self-energy and the dephasing term have to be transformed. Second, within
perturbation theory the velocity gauge induces divergent terms in the response function that in principle cancel each other, but that in practice lead to artificial divergences in the optical response\cite{PhysRevB.76.035213,PhysRevB.62.7998} due to numerical precision and incomplete
basis sets. 

The interaction Hamiltonian $U$ is evaluated in terms of
unperturbed Kohn-Sham eigenfunctions as
\be
\langle m \kk | U | n \kk \rangle = - \mathbf E(t) \langle m \kk | \mathbf r |
n\kk \rangle =  - \mathbf E(t)\,  \mathbf r_{mn,\kk},
\label{u_expantion}
\ee
where the dipole matrix elements $\mathbf r_{mn,\kk}$ , for $m \neq n$ are calculated by using the
commutation relation $i[H,\mathbf r] = \mathbf p + i[V_{\text{nl}},\mathbf
  r]$ where $V_{nl}$ is the non-local part of the Hamiltonian operator.~\cite{dipoles}

Since we are interested in calculating the dielectric properties at
zero momentum, we choose to work with an homogeneous electric
field $\mathbf E(t)$, with no space dependence
except its direction,~\cite{PhysRevB.62.7998} generated by a vector potential $\mathbf A(t)$ constant in space, 
\be
\mathbf E(t) = -\frac{1}{c} \frac{d \mathbf A(t)}{dt}.
\label{efroma}
\ee
Also in this case other choices consistent with the periodic boundary
conditions would be possible, as for example an external potential
with the cell periodicity,\cite{PhysRevLett.87.036401} or an electric
field  with a finite momentum\cite{PhysRevLett.84.1768} {\bf q} such
that $\mathbf q = \kk-\kk'$. 

Instead, the particular form of $\mathbf E(t)$ as function of time is not
specified \emph{a priori}, but given as input parameter of the simulation.
Indeed, the possibility of providing the form of the external
field as an input is one of the key strengths of the real-time approach,
potentially allowing to use the same implementation to simulate a
broad range of phenomena and of experimental techniques.  
For example, as described in Sec.~\ref{computational}
in order to calculate the linear optical susceptibility spectrum
$\chi(\omega)$ we will use a delta function ${\mathbf E(t)} = {\mathbf
  E}_0 \delta(t-t_0)$ (obtained from Eq.\eqref{efroma} with $\mathbf A(t) = A_0\Theta(t-t_0) $, where $t_0$ is the time at
which the external field is switched on). This electric field probes the system at
all frequencies with the same intensity. Also, in the other example described
in Sec.~\ref{computational},  we can use a quasi-monochromatic source
${\mathbf E(t)} = {\mathbf E}_0 \sin{\omega_0 t}\exp{(-\delta^2(t-t_0)^2/2)}$ to selectively excite
the system at a given frequency $\omega_0$.  Furthermore, two or more electric
fields can be used to simulate e.g. pump-probe, sum-of-frequency or wave-mixing
experiments.

The macroscopic quantity that is calculated at the end of the real-time
simulation is the induced polarization $\mathbf P(t)$, related to the electric displacement $\mathbf D(\mathbf r,t)$ and the electric field $\mathbf
E(\mathbf r,t)$ by the so called material equation:
\be
\mathbf D(r,t)= \epsilon_0 \mathbf E(r,t) + \mathbf P(r,t),
\ee
that stems directly from the Maxwell equations.
$\mathbf P(t)$ is obtained from $G^<$ [Eq.~\eqref{tdbse}] by,
\be
\label{pdit}
\mathbf P(t) = -\frac{1}{V}\sum_{n,m,\kk}  r^{}_{mn,\kk} G^<_{nm,\kk}(t),
\ee
and from this quantity we can obtain the optical properties of the system under study.

For instance, within linear response, the electric displacement $\mathbf D(\mathbf r,t)$ is directly proportional, in frequency space, to the electric field as $\mathbf
D(\omega) = \hat \epsilon(\omega) \epsilon_0 \mathbf E(\omega)$.  Therefore the polarization can be expressed as:
\be
\mathbf P(\omega) = \epsilon_0 (\hat \epsilon(\omega) -\hat I)  \mathbf E( \omega) 
\label{polarization}
\ee
and accordingly the optical susceptibility that describes the linear response of the system to a perturbation is $ \hat \chi(\omega) = \hat \epsilon(\omega) -\hat I$. 
Then, the optical susceptibility $\hat \chi(\omega)$ can be calculated 
by Fourier transforming the macroscopic polarization $\mathbf P(\omega)$ (or alternatively the current density $\mathbf j(\omega)$),
by means of Eq.~\eqref{polarization} as: 
\be
\label{chiomega}
\hat \chi(\omega) = \frac{\mathbf P(\omega)}{\epsilon_0 \mathbf E(\omega)}.
\ee
Note that by choosing a delta-like $\mathbf E(t)$, the Fourier transform of $\mathbf P(t)$ provides directly the full spectrum of the optical susceptibility $\hat \chi(\omega)$.
Beyond the linear regime, higher order response functions, $\hat \chi^{(2)}, \hat \chi^{(3)},\dots$ can be 
obtained (to calculate e.g. the second- or third-harmonic generation) by using a (quasi)monochromatic field as in e.g. Ref.~\onlinecite{takimoto:154114}; 
non-perturbative phenomena, such as high-harmonic generation, can be analyzed instead from the power spectrum ($|P(\omega)|^2$).

To summarize, the schematic flow of a time-dependent BSE simulation  
is shown in Fig.~\ref{fg:scheme} as has been implemented in the
development version of the{\sc Yambo} code.~\cite{yambo}  

\begin{figure}[ht]
\centering
\epsfig{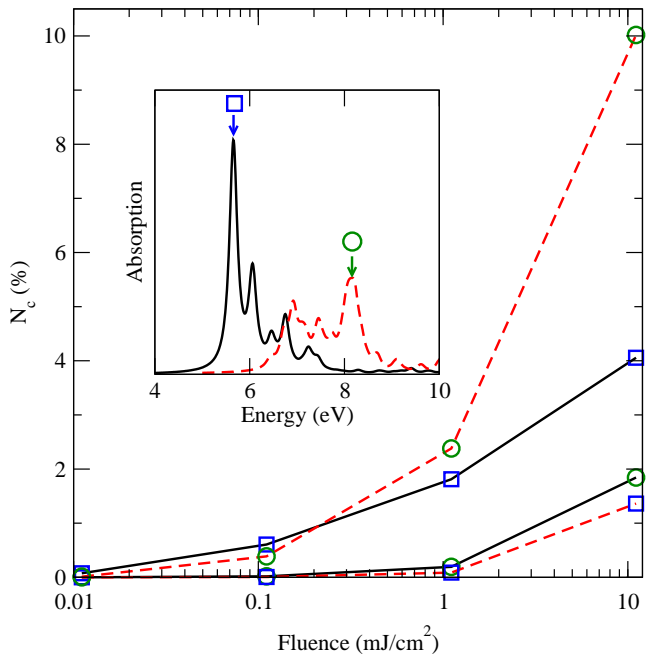}
\caption{\footnotesize{
{\it h}-BN: Percentage of valence electrons pumped to the conduction
bands ($N_c$) by a quasi-monochromatic pulse as a function of the
fluence. The pulse is centered either at 5.65\,eV (blue boxes) or 8.1\,eV (green circles)
 calculated within the TD-BSE (black line) and the td-HARTREE (red
 dashed line) approximations. In either case, each point corresponds to
 a separate simulation and the lines are drawn to help guide the eye. 
 The inset shows the absorption spectra within the TD-BSE (black line) and TD-HARTREE (red dashed line) with the arrows pointing at the pump frequencies.}} 
\label{Nc_Flu}
\end{figure}

\begin{figure*}[ht]
\centering
\epsfig{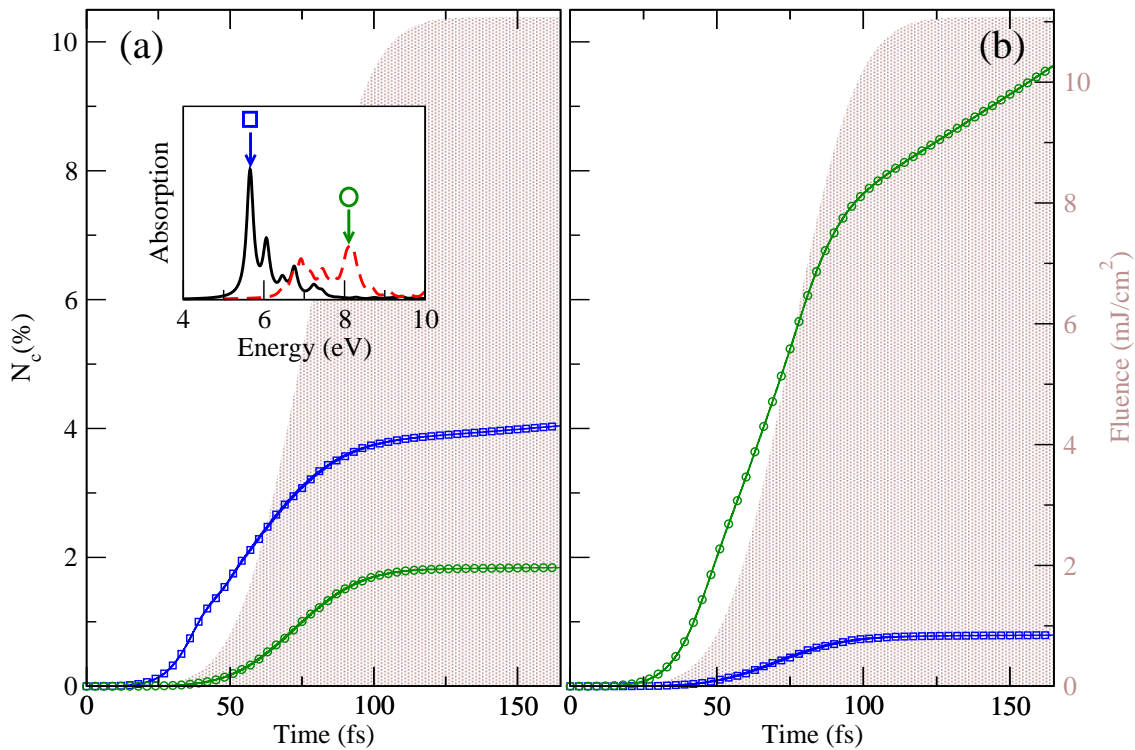}
\caption{\footnotesize{
 {\it h}-BN: Percentage as function of time of valence electrons pumped to the conduction bands ($N_c$) by a quasi-monochromatic pulse with intensity 10$^9$\,kW/cm$^2$ centered either at 5.65\,eV (blue boxes) or 8.1\,eV (green circles)
 and calculated either within the TD-BSE [{\bf(a)}] or the TD-HARTREE [{\bf(b)}]
 approximations. The brown shadow represents the fluence as function of time. 
The inset shows the absorption spectra within the TD-BSE (black line) and TD-HARTREE (red dashed line) with the arrows pointing at the pump frequencies.}} 
\label{deltaN}
\end{figure*}

\subsection{Dissipative effects}
\label{ss:stdystt}

In an excited electronic system dissipative effects are present due to inelastic electron scattering and
(quasi)-elastic scattering processes with other degrees of freedom, such as defects or phonons. Both effects contribute
to the relaxation and decay of excited electronic population  as well as of the 
decay of
phase coherence, that is to a finite dephasing rate. Our approach, Eq.~\eqref{tdbse}, does not account for dissipative
effects: on the one hand the COHSEX self-energy is real, so that the excitations lifetimes are infinite, on the other hand the electronic systems is
perfectly isolated [Eq.~\eqref{hamiltonian}], so that there is no dephasing due to interaction with other degrees of freedom.

In practical calculations then we introduce a phenomenological damping to simulate dissipative effects. We implemented two different approaches. An  \emph{a
posteriori} treatment, where at the end of the simulation (in the post-processing block of Fig.~\ref{fg:scheme}) the polarization (and the electric
field) are multiplied by a decaying exponential function, $e^{-t/\tau}$, where $\tau$ is an empirical parameter.
This parameter, that is 
 compatible with the simulation
length, effectively simulates the dephasing and introduces a Lorentzian broadening in the resulting absorption spectrum. This is in the same
spirit of the  Lorentzian broadening introduced in the linear response treatment to simulate the experimental optical spectra, and has the advantage
of producing spectra with different broadening from the same real-time simulation. Nevertheless this approach is limited to the linear response case.

In order to treat dissipative effects beyond the linear regime, an imaginary term is added to the self-energy 
in the form of an additional term ${\bf \Gamma}_{\kk}\( \GG_{\kk}^<(t)-\tilde{\GG}_{\kk}^<\)$ appearing on the r.h.s. of Eq.~\eqref{tdbse},
with:
\be
\label{eq:sciss_plus_damping}
i\[{\bf \Gamma_{\kk}}\]_{n_1,n_2} = \Gamma_{n_1\kk}^{ph}+\Gamma_{n_2\kk}^{ph}+\Gamma_{n_1\kk}^{pop}\delta_{n_1,n_2},
\ee
where $\Gamma_{n_1\kk}^{pop}$ and $\Gamma_{n_1\kk}^{ph}$ are respectively the lifetime of the perturbed electronic population and the dephasing rate, and are given as input parameters of the
simulation.

\section{Examples}
\label{computational}                                        
To illustrate and validate the time-dependent BSE approach and
our numerical implementation, 
we present two examples on {\it h}-BN. This is a wide gap insulator whose optical properties are
strongly renormalized by excitonic effects and for which all the
parameters necessary in DFT, $G_0W_0$ and response calculations,~\cite{calculations} 
are known from previous studies.~\cite{PhysRevLett.96.126104,PhysRevLett.100.189701} 

In these examples we used Eq.~\eqref{tdbse}, with and without
including the self-energy term. We refer to the
former approximation as TD-BSE, and to the latter as
TD-HARTREE. Within equilibrium MBPT these two approximations
would correspond to the BSE and RPA, and in fact they reduce to BSE and RPA
within the linear response limit (Sec.~\ref{linear_response}).   

In the first example (Fig.~\ref{hbn}), we simulated {\it h}-BN interacting with a
weak delta-like laser field.~\cite{sim1} As explained in Sec.~\ref{ss:solution}
a delta-like laser field probes all frequencies of the system and
the Fourier transform of the macroscopic polarizability provides directly the susceptibility [Eq.~\eqref{chiomega}], and thus the dielectric
constant [Eq.~\eqref{polarization}]. Since we use a weak field, we
expect negligible nonlinear effects. Then accordingly with Sec.~\ref{linear_response},  the results from
TD-BSE and TD-HARTREE can be directly compared with the BSE and RPA within
the standard {\it Ai}-MBPT approach. Indeed, in Figs.~\ref{hbn}{\bf (b)},~\ref{hbn}{\bf (d)} the imaginary part of the dielectric constant
(optical absorption) obtained by Fourier transform of the polarization
in Figs.~\ref{hbn}{\bf (a)},~\ref{hbn}{\bf (c)} is indistinguishable from that
obtained within equilibrium {\it Ai}-MBPT, validating our numerical
implementation.

In the second example (Figs.~\ref{Nc_Flu}-\ref{deltaN}) we exploit the potentiality of the TD-BSE
approach by going beyond the linear regime and using a strong
quasi-monochromatic laser field (see Sec.~\ref{ss:solution}). This field excites the system
selectively at one given frequency, moreover it is strong enough
to induce changes in the electronic population
of the system. To track these changes, during the dynamics we followed the evolution of 
$N_c(\%)$, that is the percentage of valence electrons that are pumped by the electric field in the conduction bands
(in our simulation we  have 16 valence electrons in the {\it h}-BN unit cell, since core electrons are accounted using
pseudopotentials).
The total number of valence electrons in the system is given by
the trace of $G^<$, while $N_c(t) = -i\sum_{c\kk} G^<_{cc\kk}(t)$
where $c$ labels the empty states in the unperturbed system.

We performed the simulations~\cite{sim2} for different intensities of the field (from 10$^6$\,kW/cm$^²$ to  10$^9$\,kW/cm$^²$) and for two 
values of the field frequency, 5.65\,eV and 8.1\,eV, that depending on
the level of the theory, are either at resonance or off-resonance with
the system characteristic frequencies. More precisely, within TD-BSE 5.65\,eV corresponds to the
strong excitonic feature in the absorption spectrum, while at 8.1\,eV the absorption is
negligible; conversely within RPA at 5.65\,eV the absorption is
negligible, while 8.1\,eV corresponds to the strongest feature in
the spectrum (see inset of
Figs.~\ref{Nc_Flu}-\ref{deltaN}). The results of the various
simulations are summarized in Fig.~\ref{Nc_Flu} that shows $N_c(\%)$ as
a function of the fluence---the pulse energy per unit area. For a
comparison the ablation threshold of {\it h}-BN has been determined as
78\,mJ∕/cm$^2$ in the femtosecond laser operational
mode.~\cite{10.1063/1.1787909}     

Finally, Figs.~\ref{deltaN}{\bf (a)} and~\ref{deltaN}{\bf(b)} 
show the evolution of $N_c(\%)$ during the simulation for a field
intensity of 10$^9$\,kW/cm$^²$:
one can clearly observe the enhancement in the electronic-population change
due to resonance effects. The very different picture that is obtained
within the two different approximations emphasizes the importance of
accounting for excitonic effects (also) in the strong field regime.

\section{Summary}
\label{conclusion}                                        
We presented a novel approach to the \emph{ab-initio} calculation of
optical properties in bulk materials and nano-structures that uses a
time-dependent extension of the BSE.  
The proposed approach combines the flexibility of a real-time approach
with the strength of MBPT in capturing electron-correlation.  It
allows to perform computationally feasable simulations beyond the
linear regime of e.g. second- and third-harmonic
generation, four-wave mixing, Fourier spectroscopy or pump-probe experiments.
Furthermore, being the approach based on the non-equilibrium Green's Function theory, it is possible to
include effects such as lifetimes, electron-electron
scattering\cite{bsedynamic} and electron-phonon
coupling\cite{PhysRevLett.101.106405} in a systematic way. 
Finally, we have applied the TD-BSE to the case of {\it h}-BN. First, we have calculated
the optical absorption and compared it with the results from equilibrium
{\it Ai}-MBPT validating our approach and numerical
implementation. Then, we have shown the potentialities of the TD-BSE approach beyond the linear-regime 
by calculating the change in the electronic population due to the interaction with a strong quasi-monochromatic laser field. 

\section{Acknowledgments}
\label{ackno}                                        
A.~C. acknowledges useful discussions on this work with Ilya Tokatly and Lorenzo Stella. 

The authors acknowledge funding by the European Community through e-I3 ETSF
project (Contract Number 211956). 
A.~M. acknowledges support from the HPC-Europa2 transnational programme (application N. 819).
M.~G. acknowledges support from the Funda\c{c}\~{a}o para a Ci\^{e}ncia e a Tecnologia (FCT) through the Ci\^{e}ncia 2008 programme.

This work was performed using HPC resources of the GENCI-IDRIS project No.~100063, of the CIMENT platform in Grenoble, of the CASPUR HPC
in Rome, and of the Laboratory of Advanced Computation of the University of Coimbra.
\appendix

\section{An efficient method to update the COHSEX self--energy during the time evolution}
\label{fastcohsex}
In this appendix we show how we store and update the $\Sigma^{\text{cohsex}}$ self-energy in a efficient manner. First of all we neglect the variation of the
screened interaction $W(\mathbf r,\mathbf {r'}; G^<(t))$ with respect to the $G^<(\mathbf r,\mathbf{r'},t)$ by setting to zero the functional
derivative $\partial W/\partial G$ (see Sec.~\ref{linear_response}). Within this approximation the $\Sigma^{\text{coh}}$ does not contribute to the
time evolution, therefore only $\Sigma^{\text{sex}}$ needs to be updated:
\begin{align}
\Sigma^{\text{sex}}(\mathbf r,\mathbf{r'},t) = i W(\mathbf r,\mathbf {r'})\sum_{\substack{n,n'}{\kk}} \varphi^{}_{n, \kk}(\mathbf r) 
\varphi^*_{n', \kk}(\mathbf {r'}) G^<_{n,n',\kk}(t).
\label{sex}
\end{align} 
The KBE
involves the  matrix elements $ \langle m, \kk |\Sigma^{sex} | m', \kk \rangle$:
\begin{align}
\Sigma^{\text{sex}}_{m,m',\kk}(t) = \sum_{\substack{\mathbf G,\mathbf{G'},\mathbf q \\ n,n'}} \rho^{}_{\substack{m,n \\ \mathbf{k,q}}} (\mathbf{G'})
\rho^*_{\substack{m',n' \\ \mathbf{k,q}}}(\mathbf{G}) W_{\mathbf G,\mathbf G'}(\mathbf q) G^<_{\substack{n,n' \\ \mathbf{k-q}}}(t),
\end{align}
where 
\be
\rho^{}_{\substack{m,n \\ \mathbf{k,q}}} (\mathbf{G}) = \int  \varphi^*_{m, \kk}( \mathbf r) \varphi_{n,\mathbf{k-q}}( \mathbf r)  e^{i(\mathbf G+\mathbf q) \mathbf r}.
\ee
In order to rapidly update  $\Sigma^{\text{sex}}$ after a variation of  $G^<(\mathbf r,\mathbf{r'},t)$, we store the matrix elements:
\be
M_{ \substack{m,m',n,n' \\ \mathbf q, \kk}} =  \sum_{\mathbf G,\mathbf {G'}} \rho^{}_{m,n} (\kk,\mathbf q,\mathbf{G'}) \rho^*_{m',n'}( \kk, \mathbf q,
\mathbf G) W_{\mathbf G,\mathbf G'}( \mathbf q ) ,
\ee
in such a way that $\Sigma^{\text{sex}}_{m,m'}$ can be rewritten as
\be
\Sigma^{\text{sex}}_{m,m',\kk}(t) = \sum_{\substack{n,n' \\ \mathbf{q}}} M_{\substack{m,m',n,n' \\ \mathbf q, \kk}} \cdot G^<_{\substack{n,n' \\
\mathbf{k-q}}}(t).
\ee
The $M$ matrix can be very large, but its size can be reduced by noticing that: 
(i) the matrix $M$ is Hermitian respect to the $(m,m')$ indexes; 
(ii) the number of {\bf k} and {\bf q} points is reduced by applying the operation symmetries that are left unaltered by the applied external
field; (iii) for converging optical properties only the bands close to the gap are needed (see section \ref{computational}).
As an additional numerical simplification we neglected all terms such that $M_{ \substack{m,m',n,n' \\ \mathbf q,\kk}} /\max\{ M_{ \substack{ m,m',n,n'
\\ \mathbf q , \kk}}\}<M_c$, where $M_c$ is a cutoff that, if chosen to be $M_c \simeq 5\cdot 10^{-3}$  does not appreciably affect the final results. In principle by using an auxiliary localized basis set\cite{schwegler:9708,*PhysRevB.83.115103} one can obtain a further reduction of the matrix dimensions, but in the present work we did not explore this strategy.

\addcontentsline{toc}{chapter}{Bibliography}
\bibliographystyle{apsrev4-1}
\bibliography{slowlight}

\begin{thebibliography}{10}
\bibitem{dipoles}%
  \BibitemOpen
  \bibinfo {note} {The intra-band terms, the ones for $m=n$, can be obtained by
  using the Berry curvature\cite{PhysRevB.76.035213}. However in this work we
  did not include intra-band matrix elements in our calculations. It is known
  that they do not affect optical properties of semiconductors and insulators
  within linear regime, and it is not clear yet how much they are relevant
  beyond linear optics, see ref. \onlinecite{PhysRevB.52.14636} for a
  discussion. Moreover notice that even in presence of the scissor operator,
  the optical matrix elements derived from the commutation relation are not
  affected at the lowest order by the presence of this additional non-local
  operator\cite{PhysRevB.80.155205,*PhysRevB.82.235201}.}%
  \bibAnnoteFile{Stop}{dipoles}%
\bibitem{sim2}%
  \BibitemOpen
  \bibinfo {note} {In all the simulations the time-step is 0.5\,as. As
  described in Sec.~\ref{ss:stdystt} we introduced an additional term in
  Eq.~\eqref{tdbse} to simulate dissipative effects. We chose 1\,ns for the
  lifetime of the perturbed electronic population , and 5\,fs for the dephasing
  rate.}%
  \bibAnnoteFile{Stop}{sim2}%
\end{thebibliography}
\end{document}